\documentclass[conference]{IEEEtran}
\IEEEoverridecommandlockouts
\usepackage{cite}
\usepackage{amsmath,amssymb,amsfonts}
\usepackage{algorithmic}
\usepackage{graphicx}
\usepackage{textcomp}
\usepackage{amsmath}
\usepackage{booktabs}
\usepackage{array}      
\usepackage{xcolor}
\usepackage{float}
\usepackage{subfig}
\usepackage[utf8]{inputenc}
\usepackage{CJKutf8} 

\DeclareCaptionLabelFormat{ieeefig}{\figurename\nobreakspace\thefigure.\nobreakspace\nobreakspace}
\begin{document}

\title{LEO-NA Walker Constellation Design with Bi-objective Optimisation Approaches \\
\thanks{$^{*}$Corresponding author: Junhui Zhao (junhuizhao@hotmail.com).}
}

%
%


\author{\IEEEauthorblockN{1\textsuperscript{st} Chun Zhang}
	\IEEEauthorblockA{\textit{School of Electronic and Information } \\
		\textit{Engineering, Beijing Jiaotong University}\\
		Beijing 100044, China \\
		zc17636268079@163.com}
	\and
	\IEEEauthorblockN{2\textsuperscript{nd} Junhui Zhao$^{*}$}
	\IEEEauthorblockA{\textit{School of Electronic and Information } \\
		\textit{Engineering, Beijing Jiaotong University}\\
		Beijing 100044, China \\
		junhuizhao@hotmail.com}
	\and
	\IEEEauthorblockN{3\textsuperscript{rd} Zehan Liu}
	\IEEEauthorblockA{\textit{School of Electronic and Information } \\
		\textit{Engineering, Beijing Jiaotong University}\\
		Beijing 100044, China \\
		23140010@bjtu.edu.cn}
	\and
	\IEEEauthorblockN{4\textsuperscript{th} Xinpeng Liu}
	\IEEEauthorblockA{\textit{School of Electronic and Information } \\
		\textit{Engineering, Beijing Jiaotong University}\\
		Beijing 100044, China \\
		24125047@bjtu.edu.cn}
	\and
	\IEEEauthorblockN{5\textsuperscript{th} Shaoyi Xu}
	\IEEEauthorblockA{\textit{School of Electronic and Information } \\
		\textit{Engineering, Beijing Jiaotong University}\\
		Beijing 100044, China \\
		shyxu@bjtu.edu.cn}
	\and
	\IEEEauthorblockN{6\textsuperscript{th} Xiaoming Wang}
	\IEEEauthorblockA{\textit{School of Communication } \\
		\textit{and Information Engineering }\\
		\textit{Nanjing University of}\\
		\textit{Posts and Telecommunications}\\
		Nanjing 210003, China \\
		xmwang@njupt.edu.cn}
}

\maketitle

\begin{abstract}
Low Earth Orbit (LEO) constellation design for navigation augmentation (NA) has attracted increasing attention in navigation satellite system studies, yet balancing navigation performance and deployment cost remains a fundamental challenge. To address this issue, this paper proposes a bi-objective optimization framework for LEO Walker constellation design. The problem is formulated as a bi-objective optimization model with constellation cost and positioning accuracy as objectives. In the formulation, PDOP tail risk and satellite visibility are incorporated into the objective formulation to better characterize navigation performance. The Pareto-optimal solution set is obtained using the Non-dominated Sorting Genetic Algorithm II (NSGA-II). Simulation results show that, under the same satellite deployment cost, the proposed LEO-NA Walker constellation improves the average number of visible satellites by 42.5\% and 24.4\%, and reduces the mean PDOP by 18.9\% and 10.5\% compared with representative Polar and optimized-LFC constellations, respectively, thereby enhancing service continuity and resource utilization efficiency. These results provide useful guidance for the design and deployment of LEO-NA constellations.
\end{abstract}

\begin{IEEEkeywords}
LEO Walker constellation design, navigation performance, satellite deployment cost, bi-objective optimization.
\end{IEEEkeywords}
\section{Introduction}
Low Earth Orbit (LEO) constellations have attracted considerable attention for navigation augmentation (NA) due to their low propagation loss, fast-varying geometry and wide-area coverage capability \cite{10437778,10681995}. Compared to medium Earth orbit (MEO) and geostationary orbit (GEO) satellites, LEO constellations can significantly improve satellite geometry, thereby enhancing positioning accuracy and service performance. As a result, they have become an important research direction for next-generation global navigation satellite systems (GNSS) \cite{11162074,11008672,10320306}. This has motivated extensive studies on the effects of various LEO constellation configurations on GNSS augmentation performance.

Typical LEO constellation architectures include Polar, Iridium-like, Lattice Flower and Walker constellations, which differ significantly in orbital distribution and spatial geometry, leading to distinct impacts on GNSS availability and positioning accuracy \cite{11049853,10980228,10992253}. Among them, Polar and Iridium-like constellations are mainly used to assess high-latitude coverage and uniformity, while Teledesic constellations are primarily analyzed for large-scale global coverage. In contrast, Walker constellations, due to their regular orbital structure and favorable geometric uniformity, are widely used in the design and optimization of navigation augmentation systems \cite{guan2020optimal,11157880,11159552}.

Walker constellations provide favorable coverage uniformity and high design flexibility. Its design typically requires a trade-off between navigation performance and deployment cost. On one hand, most existing studies focus solely on navigation performance and use the Position Dilution of Precision (PDOP) as the evaluation metric. However, under conditions of poor satellite visibility, the tail risk of PDOP can significantly degrade GNSS positioning performance. In \cite{9262624} and \cite{wang2021design}, single-objective optimization methods, such as Genetic Algorithm (GA) and Particle Swarm Optimization (PSO), are applied to optimize the mean Position Dilution of Precision (PDOP). On the other hand, only a few studies consider both navigation performance and deployment cost simultaneously. In these designs, the problem is formulated as a multi-objective optimization model, in which navigation performance and deployment cost are typically quantified by the mean PDOP and the total number of satellites, respectively. Evolutionary algorithms such as Non-dominated Sorting Genetic Algorithm II (NSGA-II), Multi-Objective Particle Swarm Optimization (MOPSO) and the Multi-Objective Evolutionary Algorithm based on Decomposition (MOEA/D) are commonly used to produce Pareto-optimal solutions in multi-objective optimization \cite{yi2021leo,melaku2023optimization,11299844}.

Despite these efforts, existing methods still primarily rely on mean PDOP-based performance evaluation and largely overlook the tail risk of PDOP as well as satellite visibility. As a result, they are unable to achieve a well-balanced trade-off between performance stability and system cost, which limits their applicability to systematic constellation design. To address these issues, this paper proposes a bi-objective optimization framework for LEO Walker constellation design, where constellation deployment cost and navigation performance are jointly optimized. The NSGA-II is adopted to obtain the Pareto-optimal solution set. Simulation results demonstrate that the proposed method can significantly enhance navigation augmentation performance under reduced constellation deployment cost, compared with existing constellation configurations and objective formulations.

\section{Problem Formulation of LEO-NA Walker Constellation Optimization}

In this section, we formulate a bi-objective optimization model for LEO-NA Walker constellation.

\subsection{Walker Constellation Parameterization}
The geometry of a standard Walker constellation can be uniquely characterized by the parameter vector
\begin{equation}
{\bf{x}}= [P,S,i,h,\Delta \phi ,F]^T\label{eq1}
\end{equation}
where $P$ denotes the number of orbital planes, $S$ denotes the number of satellites per plane, $i$ is the orbital inclination, $h$ is the orbital altitude, ${\Delta \phi}={{\rm{2}}\pi}/S$ represents the intra-plane angular separation and $F$ denotes the inter-plane phasing parameter.

Each satellite follows a Keplerian orbit, which can be fully described by six orbital elements: the semi-major axis $a$, eccentricity $e$, inclination $i$, argument of perigee $\omega_{k,j}$, right ascension of the ascending node $\Omega_{k,j}$  and mean anomaly $M_{k,j}$. The right ascension of the ascending node for the $j$-th satellite in the $k$-th plane is given by
\begin{equation}
	\Omega_kj = \frac{2\pi (k-1)}{P}\label{eq2}
\end{equation}	
and its mean anomaly is
\begin{equation}
	M_{k,j}=\frac{2\pi (j-1)}{S}+\frac{2\pi F (k-1)}{PS}.\label{eq3}
\end{equation}

To ensure the practical feasibility of the constellation deployment while achieving a balanced trade-off among coverage performance, communication latency and deployment cost, the optimization variables are constrained as follows:
\begin{equation}
	400km\leq h \leq 1000km,\label{eq4}
\end{equation}
\begin{equation}
	40^\circ \leq i \leq 60^\circ,\label{eq5}
\end{equation}
\begin{equation}
 11 \le P \le 15,11 \le S \le 15,
P,S \in \mathbb{Z}^+.\label{eq6}
\end{equation}

The orbital constraints in (4)–(6) are defined for a regional NA scenario targeting mid-latitude coverage (e.g., China and surrounding regions). The altitude range (400–1000 km) and inclination range (40°–60°) are selected to balance coverage performance and orbital feasibility, while the number of orbital planes and satellites per plane (11–15) is constrained to reduce optimization complexity. These settings limit the search space and are not intended for large-scale broadband LEO constellations (e.g., Starlink-type systems) with substantially different design objectives and scales.

\subsection{NA Performance Metrics}
Consider a joint-observation NA positioning system consisting of GNSS (e.g., Beidou-3) and enhanced LEO satellite, and establish corresponding NA performance evaluation metrics. In this system, the clock offsets of heterogeneous systems are modeled as state parameters within a unified framework, and jointly estimated using least-squares or filtering-based methods. The primary objective is to improve positioning accuracy, which is evaluated in terms of the Position Dilution of Precision (PDOP) metric, as defined in (7).
\begin{equation}
	\mathrm{PDOP}(g,t)
	=
	\sqrt{
		\operatorname{tr}
		\left(
		\left[((\mathbf{G}(g,t))^{T}\mathbf{G}(g,t))^{-1}\right]_{1:3,1:3}
		\right)
	}\label{eq7}
\end{equation}
where $\mathbf{G}$ is the observation geometry matrix, $tr\left(\bullet\right)$ denotes the matrix trace operator and ${\left[  \cdot  \right]_{{\rm{1:3,1:3}}}}$ extracts the position-related submatrix.

Under adverse geometric conditions, the tail risk of PDOP may significantly degrade positioning reliability. To jointly account for positioning accuracy and reliability, we define the following composite metric:
\begin{equation}
	J_{pos} = \mathbb{E}_{g,t}(PDOP(g,t)) + \lambda \mathrm{CVaR}_{95\%}(PDOP(g,t))\label{eq8}
\end{equation}
where $\lambda$ is a weighting coefficient and ${CVaR}_{95\% }\left(  \cdot  \right)$ denotes the $95$\% Conditional Value-at-Risk (CVaR).

Another key factor is the constellation deployment cost. Since a larger constellation generally provides more visible satellites at the expense of increased deployment cost, we introduce a resource-efficiency metric to characterize the deployment cost per visible satellite, defined as
\begin{equation}
	J_{cost} = \frac{P \times S}{\bar{N}_{{visible}}}.\label{eq9}
\end{equation}
where 
\begin{equation}
	\bar{N}_{{visible}} = \frac{1}{|\mathbf{G}||\mathbf{T}|}\sum_{g\in \mathbf{G}}\sum_{t\in \mathbf{T}}N_{{visible}}(g,t),\label{eq10}
\end{equation}
\begin{equation}
	N_{{visible}}(g,t)=\sum_{i=1}^{N_{\mathrm{sat}}}I_i(g,t),\label{eq11}
\end{equation}
\begin{equation}
	I_i(g,t)=
	\begin{cases}
		1, & \text{if satellite } i \text{ is visible at} \ (g,t),\\
		0, & otherwise
	\end{cases},\label{eq12}
\end{equation}
$\left| \mathbf{G} \right|$ denotes the total number of grid points in the coverage region, $\left| \mathbf{T} \right|$ is the total number of time samples. $N_{\text{visible}}(g,t)$ represents the number of visible satellites at grid point $g$ and time $t$, ${\bar N_{visible}}$ represents the average number of visible satellites. ${N_{sat}} = P \times S$ denotes the total number of satellites in the constellation.
\subsection{Bi-Objective Optimization Problem Formulation}
Based on the parameterized Walker constellation model and the defined navigation performance metrics, the LEO-NA Walker constellation design can be formulated as a bi-objective optimization problem. The design aims to maximize navigation performance while minimizing constellation deployment cost under limited resources. Formally, the objectives are:
\begin{equation}
	\begin{aligned}
        \mathop {\min }\limits_{\bf{x}}& \left( {{J_{pos}},{J_{\cos t}}} \right)\\
		\text{s.t.} &\quad \mathbf{\mathbf{x}} \in \Omega
	\end{aligned}\label{eq13}
\end{equation}
where $\Omega$ denotes the feasible parameter space defined by (4)-(6).

\section{Bi-Objective Optimization for LEO-NA Walker Constellation}
The LEO-NA Walker constellation design has been formulated as a bi-objective optimization problem. To search for the Pareto-optimal solution set, the NSGA-II is employed.
\subsection{The Pareto-optimal Solution for LEO-NA Constellation}
As illustrated in (8) and (9), the objectives $J_{pos}$ and $J_{cost}$ are inherently conflicting. Consequently, the bi-objective optimization problem in (13) does not admit a single solution that simultaneously optimizes both objectives. Instead, its solution is described by a Pareto-optimal set, which represents the trade-off frontier between navigation performance and constellation deployment cost.

Let $\mathbf{x}_1$ and $\mathbf{x}_2$ denote two feasible solutions in the feasible region $\Omega$. Solution $\mathbf{x}_1$ is said to dominate solution $\mathbf{x}_2$, if the following conditions are satisfied
\begin{equation}
J_i(\mathbf{x}_1)\le J_i(\mathbf{x}_2),\forall i=1,\ldots M \label{eq14}
\end{equation}
\begin{equation}
J_j(\mathbf{x}_1)<J_j(\mathbf{x}_2), \exists j=1,\ldots M. \label{eq15}
\end{equation}

Based on this dominance relation, the Pareto-optimal solution set of the problem in (13) is defined as
\begin{equation}
	\mathcal{P}=
	\left\{
	\mathbf{x}\in\Omega
	\;\middle|\;
	\nexists\,\mathbf{y}\in\Omega,
	\mathbf{y}\prec\mathbf{x}
	\right\}\label{eq16}
\end{equation}
where $M$ denotes the number of objectives, $\mathbf{y}\prec\mathbf{x}$ indicates that solution$\mathbf{y}$ Pareto-dominates solution $\mathbf{x}$. ${\cal P}$ represents all non-dominated solutions.
\subsection{Optimizing LEO-NA constellation with NSGA-II algorithm}
Since NSGA-II can effectively maintain population diversity while ensuring computational efficiency, it is adopted to solve the bi-objective optimization problem in (13). The overall algorithm framework is illustrated in Fig. 1.

(1)Initialization and Fitness Evaluation: An initial population of Walker constellation  ${{\bf{P}}_0} = \left\{ {{{\bf{x}}_\zeta }} \right\}_{\zeta  = 1}^{{N_p}}$ of size $N_p$ is randomly generated. For each individual, the corresponding satellite orbital states are propagated. Based on the generated constellation geometry, the average number of visible satellites $\bar{N}_{{visible}}$ and PDOP are evaluated over all grid points $g$ and time samples $t$. Accordingly, the two objective functions, i.e., the navigation performance metric $J_{pos}$ and the deployment cost metric $J_{cost}$ are computed for each individual in the population.

\begin{figure}[htbp]
	\centerline{\includegraphics[width=\columnwidth, trim={20pt 20pt 20pt 20pt}, clip]{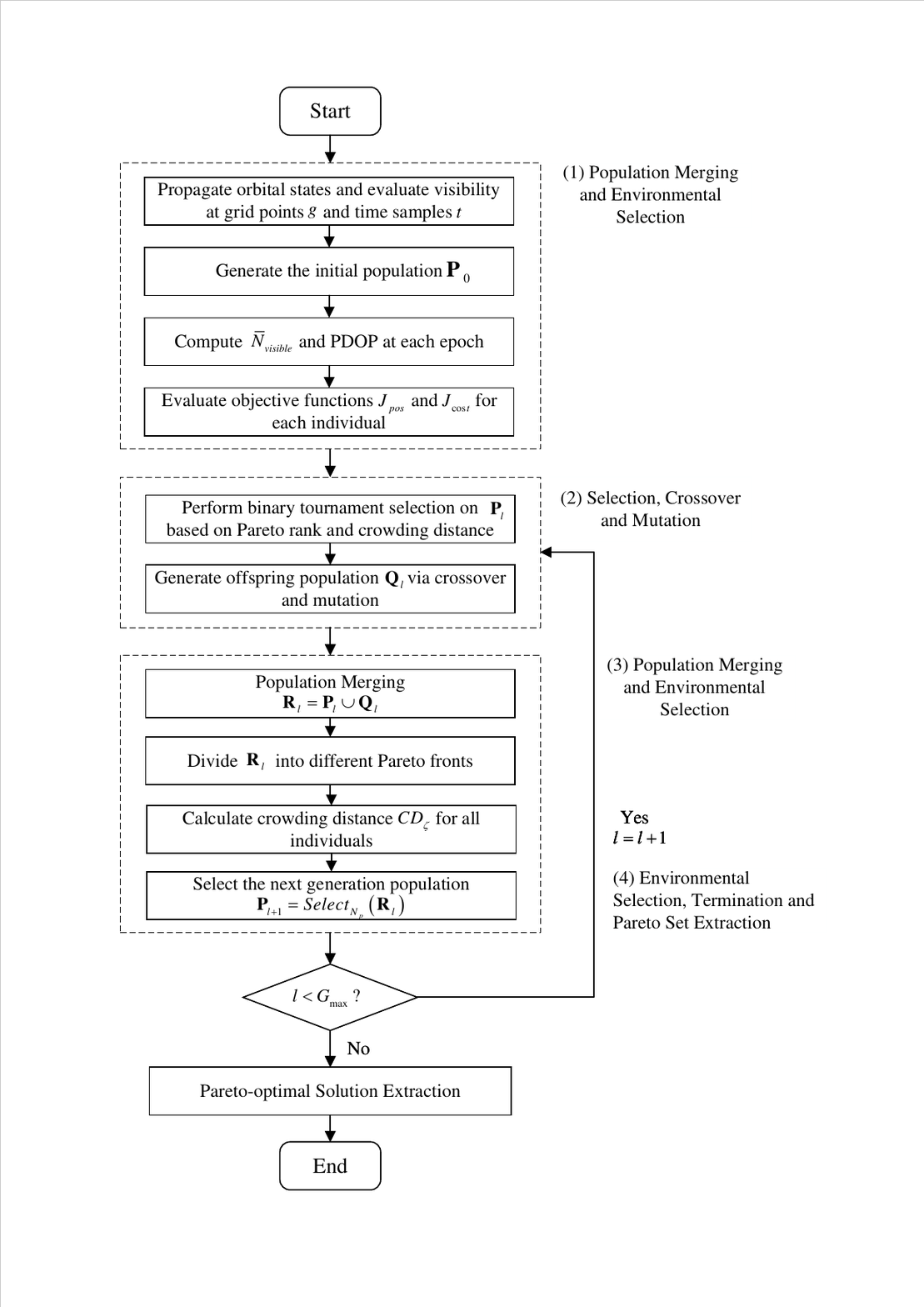}}
	\caption{The flowchart of the NSGA-II algorithm.}
	\label{fig1}
\end{figure}

(2)Selection, Crossover and Mutation: Based on the current population $\mathbf{P}_l$, parent individuals are selected using binary tournament selection according to Pareto rank and crowding distance in (17). Offspring population $\mathbf{Q}_l$ is generated through crossover and mutation operations:
\begin{equation}
	CD_\zeta = \sum_{m=1}^{2} \frac{J_m^{\zeta+1} - J_m^{\zeta-1}}{J_m^{\text{max}} - J_m^{\text{min}}}, \,\zeta  = 1, \ldots ,{N_p}.\label{eq17}
\end{equation}
\begin{equation}
	\mathbf{Q}_l = \text{Mutation}\big(\text{Crossover}(\mathbf{P}_l)\big).\label{eq18}
\end{equation}

where $J_m$ denotes the $m$-th objective value, $J_m^{\max}$ and $J_m^{\min}$ respectively represent the maximum and minimum values of the $m$-th objective within the current population.

(3)Population Merging and Non-dominated Sorting: The parent and offspring populations are merged to form a combined population:
\begin{equation}
	\mathbf{R}_l = \mathbf{P}_l \cup \mathbf{Q}_l.\label{eq19}
\end{equation}
Subsequently, fast non-dominated sorting is performed on $\mathbf{Q}_l$, and the crowding distance of each individual is computed using (17).

(4)Environmental Selection, Termination and Pareto Set Extraction: The next generation population is selected from $\mathbf{R}_l$ based on Pareto rank and crowding distance:
\begin{equation}
	\mathbf{P}_{l+1} = \text{Select}_{N_p}\big(\mathbf{R}_l\big).\label{eq20}
\end{equation}
Steps (2)–(4) are repeated until the maximum generation $G_{\text{max}}$ is reached. The final non-dominated set is obtained as
\begin{equation}
\mathcal{F}^* = \left\{ \mathbf{x}_1^*, \mathbf{x}_2^*, \dots, \mathbf{x}_{N_p}^* \right\}.\label{eq21}
\end{equation}
This set provides the trade-off solutions between constellation deployment cost and navigation performance.

\section{RESULTS AND ANALYSES}

This section evaluates the performance of the optimized LEO-NA Walker constellation and compares it with alternative constellation schemes.

\subsection{Parameter Setting}
In this study, LEO satellites are introduced as navigation augmentation sources and jointly observed and integrated with the Beidou-3 system. The system parameters of the Beidou-3 constellation are summarized in Table I. Orbit propagation is modeled using a $J2$ perturbation model and the inter-plane phasing factor is set to $F=1$. All simulations are conducted in a MATLAB–STK co-simulation environment, with a simulation duration of $24 h$ and a sampling interval of $60 s$. The CVaR weighting coefficient is set to $\lambda=0.5$, which empirically balances mean performance and tail-risk sensitivity. The coverage region spans latitudes from $3^\circ$N to $54^\circ$N and longitudes from $73^\circ$E to $136^\circ$E, and is discretized into a $5^\circ \times 5^\circ$ grid to cover the main service region of the Beidou-3 navigation system. 
\begin{table}[htbp]
	\centering
	\caption{The parameter of Beidou-3} 
	\label{tab:gps-glonass}
	\begin{tabular}{@{}>{\raggedright\arraybackslash}p{4.2cm}lll@{}}
		\toprule
		\textbf{System orbit}
		& \multicolumn{1}{c}{\textbf{Beidou-3}}\\
		\cmidrule(lr){2-4} 
		& \textbf{MEO} & \textbf{IGSO} & \textbf{GEO} \\
		\midrule
		Number of satellites        & 24               & 3       &3 \\
		Constellation pattern       & Walker(24/3/1)   & /        &/    \\
		Inclination ($^{\circ}$)    & 56               & 55      &0  \\
		Altitude (km)               & 21528            & 35786   &35786   \\
		\bottomrule    
	\end{tabular}
\end{table}

To evaluate the effectiveness of the proposed design, a set of benchmark constellations featuring diverse orbital geometries and different optimization methods are selected. Polar and Iridium-like constellations provide classical high-inclination and uniform coverage designs, respectively, while the MOPSO-optimized Lattice Flower Constellation (optimized-LFC) in [14] represents a recent optimization-based lattice constellation design. These baselines can comprehensively evaluate the proposed Walker constellation optimization framework under diverse configurations. The NSGA-II is configured with a population size $N_p=30$, the number of generations $G_{max}=30$, a crossover probability of $0.8$ and the mutation probability of $0.1$. 

Under the above settings, the Pareto-optimal solution sets of the different constellation designs are presented in Table II.
\begin{table}[htbp]
	\centering
	\caption{Pareto-optimal solutions for different constellation designs}
	\label{tab:pareto_solutions}
	
	\small
	\setlength{\tabcolsep}{3pt}
	
	\begin{tabular}{l c c c c c c}
		\hline
		\textbf{Constellation} & \textbf{No.} & $\mathbf{P} \mathbf{\times} \mathbf{S}$ & $\mathbf{\bar{N}}_{visible}$ & ${\mathbf{E}}(PDOP)$ & $\mathbf{i}(^\circ)$ & $\mathbf{h}$ (km) \\
		\hline
		
		Proposed & 1 & 195 & 8.73 & 1.1439 & 56.9 & 925 \\
		Proposed & 2 & 182 & 8.15 & 1.1630 & 55.0 & 925 \\
		Proposed & 3 & 210 & 9.31 & 1.0894 & 54.4 & 916 \\
		
		\hline
		Polar & 1 & 195 & 6.13 & 1.3519 & 90.0 & 925 \\
		Polar & 2 & 182 & 5.72 & 1.4339 & 90.0 & 925 \\
		Polar & 3 & 210 & 6.52 & 1.3208 & 90.0 & 916 \\
		
		\hline
		Iridium-like & 1 & 195 & 6.15 & 1.2986 & 85.6 & 925 \\
		Iridium-like & 2 & 182 & 5.74 & 1.3551 & 86.4 & 925 \\
		Iridium-like & 3 & 210 & 6.55 & 1.2791 & 83.2 & 916 \\
		
		\hline
		optimized-LFC  & 1 & 169 & 6.08 & 1.3636 & 83.9 & 1000 \\
		optimized-LFC  & 2 & 182 & 6.55 & 1.2987 & 78.2 & 1000 \\
		
		\hline
	\end{tabular}
\end{table}

%
%
%
%
%
%
\subsection{Results Analysis}
(1) Pareto-optimal Set and Decision Analysis

Fig. 2 gives the Pareto-optimal solutions and fuzzy decision results under different constellation designs. As shown in Fig. 2(a), the proposed constellation design method yields a Pareto-optimal solution set located in the lower-left region of all compared schemes, indicating superior performance in both optimization objectives.

\begin{figure*}[!t]
	\centering
	\subfloat[]{
		\includegraphics[width=0.48\textwidth,height=6cm,keepaspectratio]{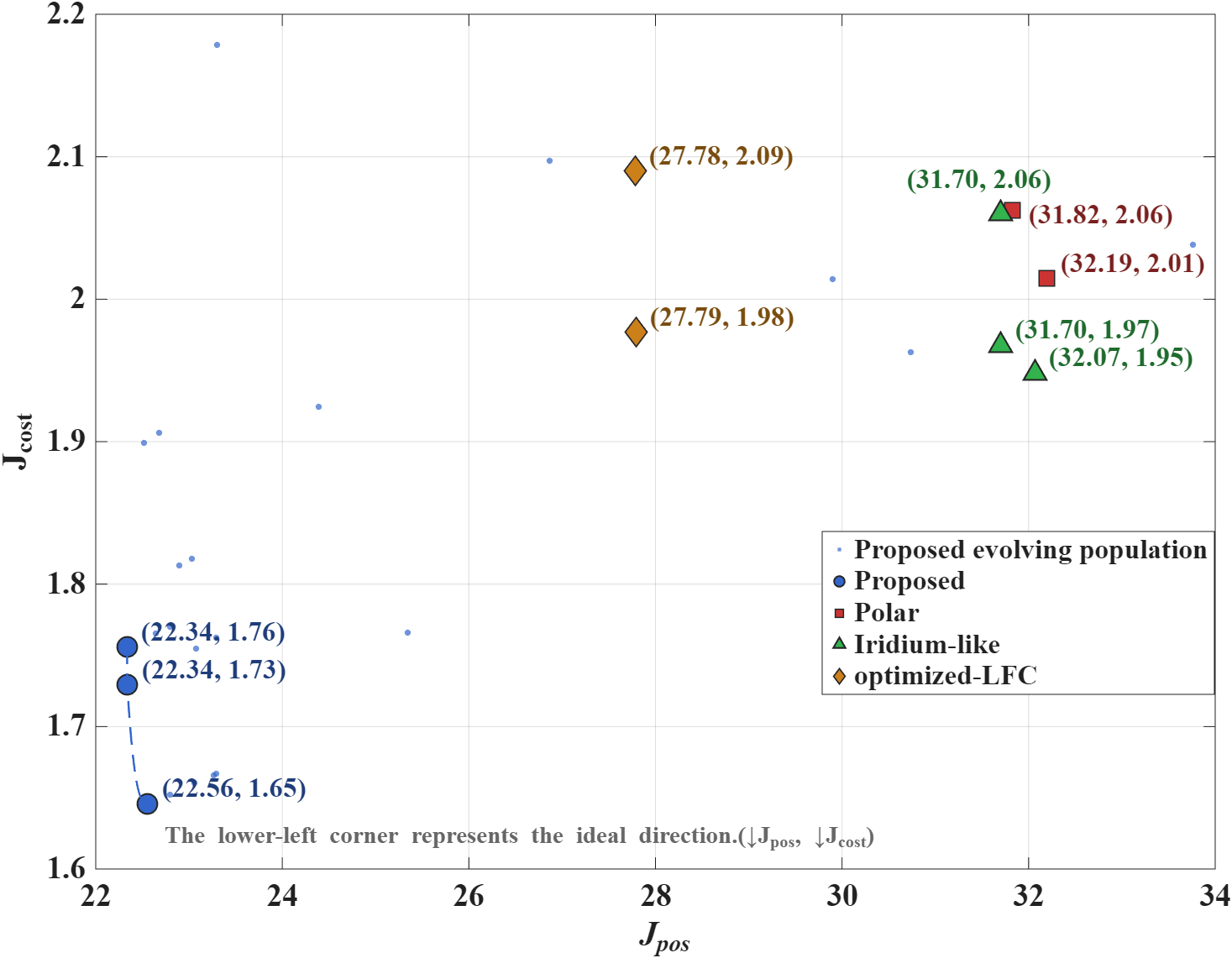}
	}
	\hfill
	\subfloat[]{
		\includegraphics[width=0.48\textwidth,height=6cm,keepaspectratio]{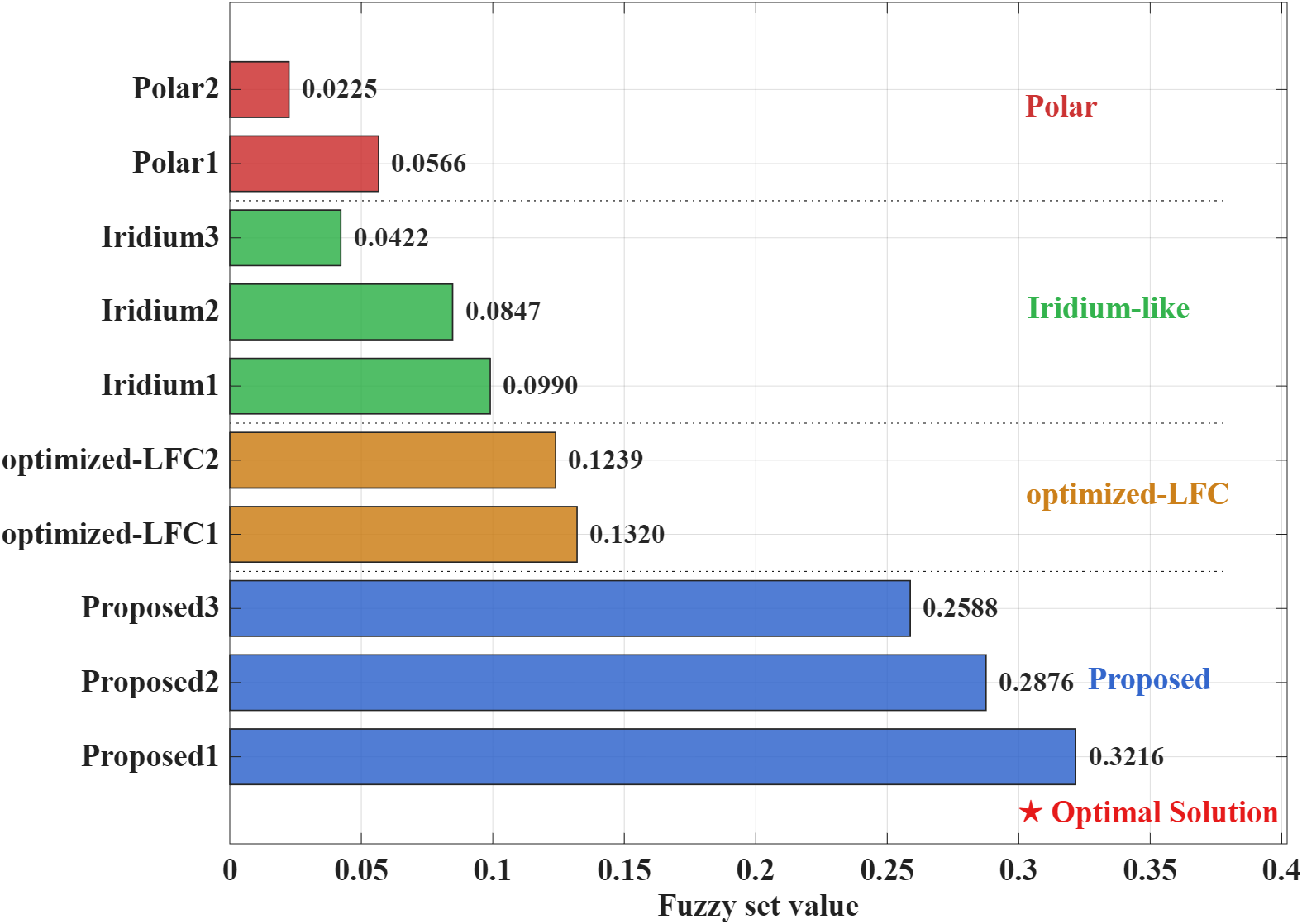}
	}
	\caption{Pareto-optimal solutions and fuzzy decision results of different constellation designs}
	\label{fig2}
\end{figure*}
\begin{figure*}[!t]
	\centering
	\subfloat[]{
		\includegraphics[width=0.425\textwidth,height=6cm]{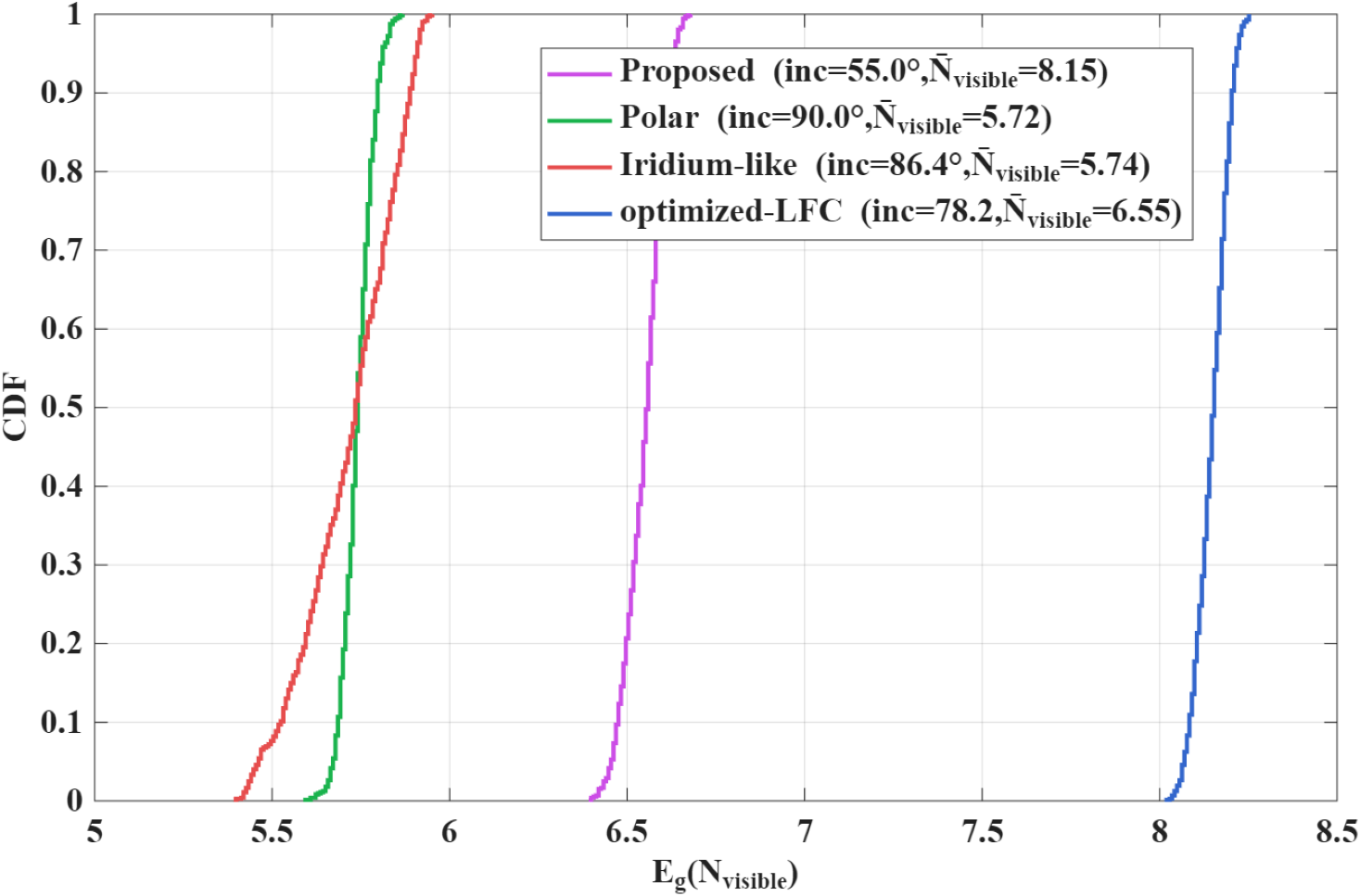}
	}
	\hfill
	\subfloat[]{
		\includegraphics[width=0.425\textwidth,height=6cm]{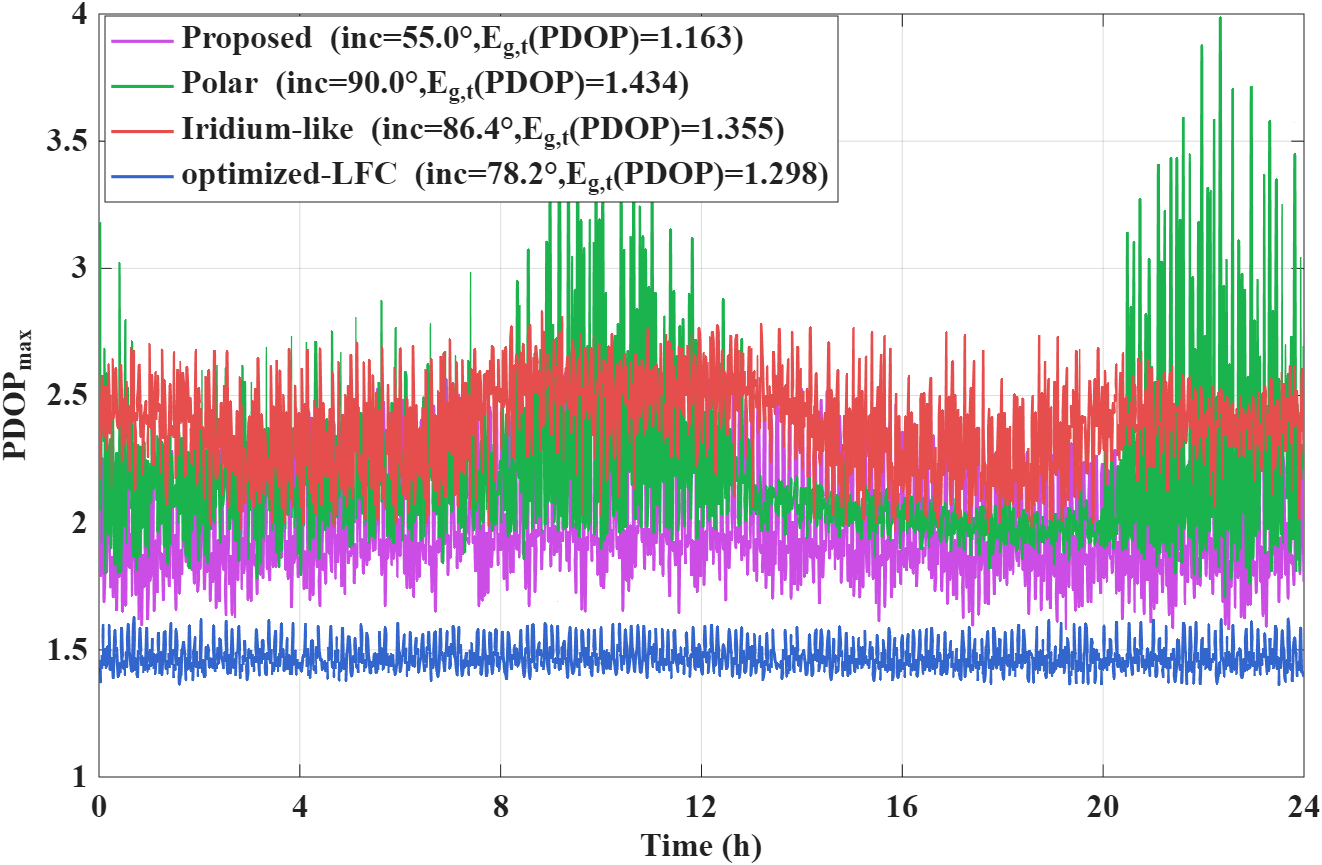}
	}
	\caption{(a) CDF of $E_g(N_{visible})$ for different constellation designs. (b) $PDOP_{max}$ for different constellation designs. }
	\label{fig3}
\end{figure*}
\begin{figure*}[!t]
	\centering
	\subfloat[]{
		\includegraphics[width=0.425\textwidth,height=6cm]{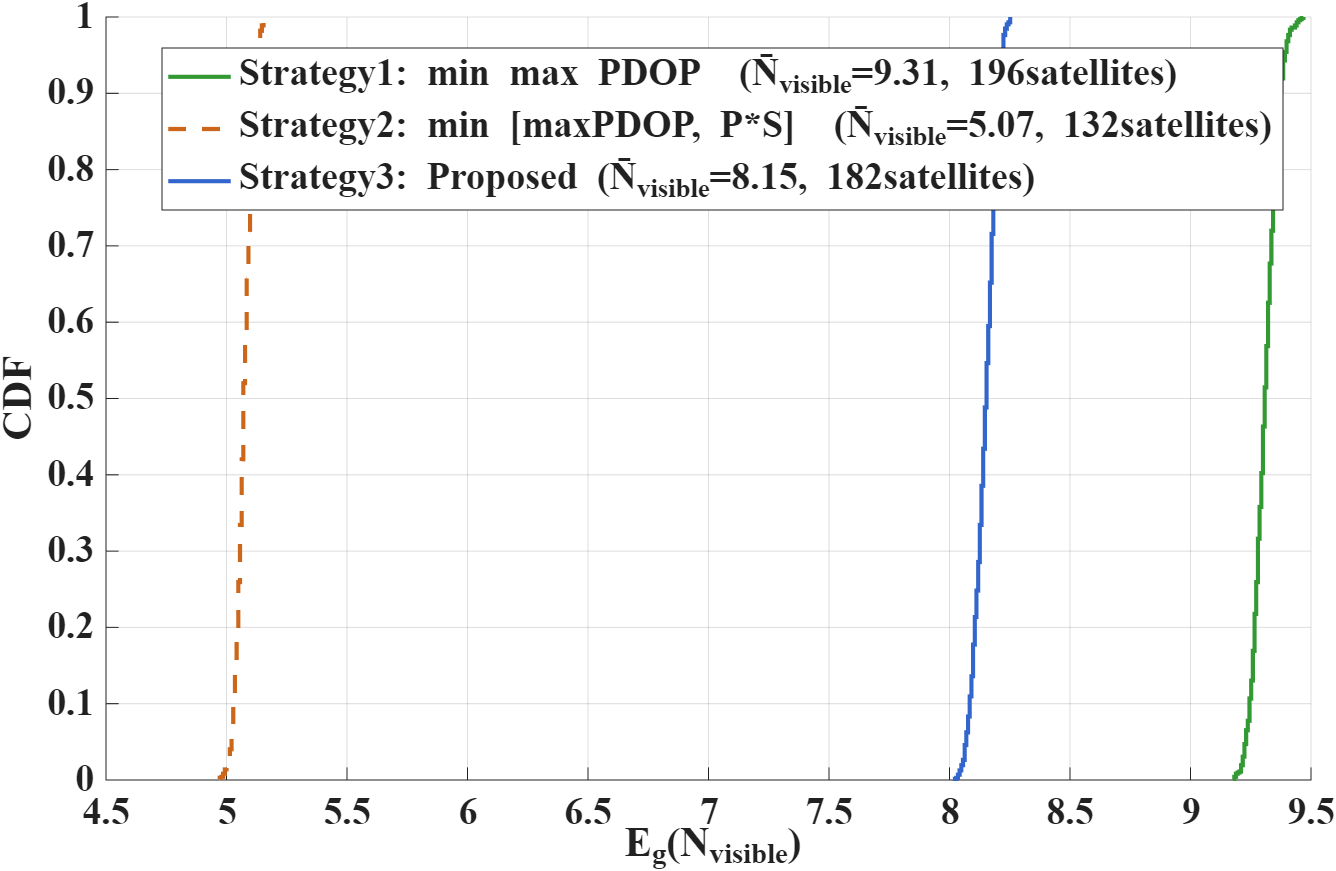}
	}
	\hfill
	\subfloat[]{
		\includegraphics[width=0.425\textwidth,height=6cm]{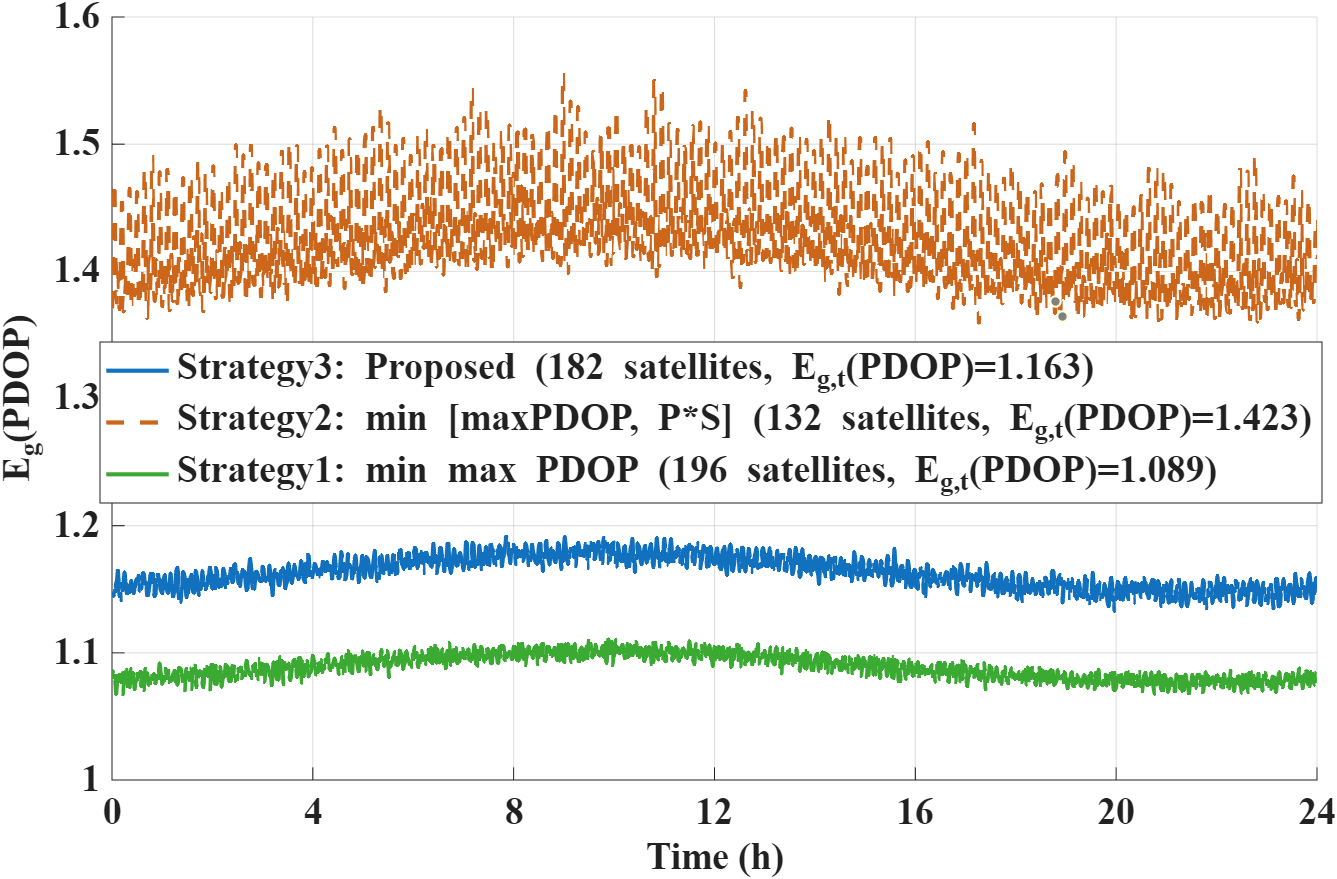}
	}
	\caption{(a) CDF of $E_g(N_{visible})$ for different optimization strategies. (b) $E_{g}(PDOP)$ for different optimization strategies.}
	\label{fig4}
\end{figure*}
To further select a single trade-off solution from the obtained Pareto-optimal sets, a fuzzy-set-based decision method \cite{bellman1970decision} is employed, as illustrated in Fig. 2(b). For a fair comparison, all schemes employ identical objective functions and a unified normalization strategy, with the values of fuzzy set computed over the combined Pareto-optimal set to ensure consistency and comparability of the results. From Fig. 2(b), it can be seen that the proposed constellation design achieves the highest fuzzy membership value, indicating an optimal trade-off between navigation performance and constellation resource cost.


(2) Comparison of Navigation Performance

We evaluate the navigation performance of different constellation designs from two aspects: satellite visibility and geometric positioning accuracy stability. Specifically, satellite visibility is characterized by the average number of visible satellites ${\bar N_{visible}}$ and the spatial average of visible satellites $E_g(N_{visible})$, which reflect the overall system availability and spatial distribution characteristics. Geometric positioning accuracy stability is evaluated using the maximum Position Dilution of Precision $PDOP_{max}$, the spatio-temporal average Position Dilution of Precision $E_{g,t}(PDOP)$ and the spatially averaged Position Dilution of Precision $E_g(PDOP)$, which characterize the worst-case positioning performance, average positioning performance and spatial stability, respectively, as shown in (23)-(24). For a fair comparison, all schemes are evaluated under constellation configurations with the same total number of satellites as listed in Table II.
\begin{equation}
E_g(N_{visible})
= \frac{1}{|\mathbf{G}|}
\sum_{g \in \mathbf{G}} N_{\mathrm{visible}}(g,t).\label{eq22}
\end{equation}
\begin{equation}
	PDO{P_{\max }} = \mathop {\max }\limits_{g,t} PDOP\left( {g,t} \right).\label{eq23}
\end{equation}
\begin{equation}
      E_g(PDOP) = \frac{1}{|\mathbf{G}|} \sum_{g \in \mathbf{G}} PDOP(g,t).\label{eq24}
\end{equation}

Fig. 3 presents the navigation performance of different constellation designs. It can be observed that, in terms of satellite visibility, the Cumulative Distribution Function (CDF) curve of the spatially averaged number of visible satellites for the proposed constellation is shifted toward higher values, indicating that a greater number of visible satellites can be maintained at the same spatial locations. Meanwhile, the proposed constellation achieves an average of 8.15 visible satellites, representing improvements of 24.4\%, 42.0\%, and 42.5\% over the optimized-LFC, Iridium-like and Polar constellations, respectively. In terms of geometric positioning accuracy stability, the proposed constellation consistently has the lowest $PDOP_{max}$ with relatively small fluctuations, demonstrating improved geometric configuration stability and superior positioning performance.

Fig. 4 investigates the impact of different optimization strategies on navigation performance. The comparison considers three optimization strategies: single-objective optimization that only minimizes the maximum PDOP, bi-objective optimization that jointly considers the total number of satellites and the maximum PDOP, and the proposed optimization strategy.
 
 As shown in Fig. 4, the single-objective strategy achieves better performance in terms of $E_g(PDOP)$ and the number of visible satellites. However, without constraints on the constellation scale, the design requires a large number of satellites, increasing deployment cost and limiting practical feasibility. In contrast, the bi-objective strategy effectively reduces the constellation size and achieves the lowest deployment cost, but its navigation performance degrades. Compared with these two strategies, the proposed method achieves a better trade-off between navigation performance and constellation deployment cost, making it more suitable for practical constellation design.



\section{CONCLUSION}

This paper proposes a bi-objective NSGA-II optimization framework for LEO-NA Walker constellation design, aiming to jointly optimize navigation performance and deployment cost. The formulation incorporates PDOP tail risk and satellite visibility into the objective functions, and derives the Pareto-optimal solution set using NSGA-II. Compared with existing constellation design methods, the proposed approach demonstrates superior navigation performance.

The proposed framework offers a practical blueprint for deploying regional navigation overlay systems, potentially benefiting latency-sensitive applications such as autonomous driving and emergency response in areas with obstructed sky visibility. Future work will extend this framework to multi-layer hybrid constellations integrating MEO and LEO assets. Additionally, dynamic optimization strategies under real-time orbital perturbations will be investigated, along with the impact of inter-satellite link (ISL) latency on time-to-first-fix (TTFF) in dense urban canyon environments.
%
%
%
%

\bibliographystyle{IEEEtran}

\bibliography{IEEEabrv,ref}

\end{document}